\documentstyle[aps,prl,epsfig,multicol]{revtex}

\begin{document}
\title{Substitution for Cu in the electron-doped infinite-layer superconductor Sr$%
_{0.9}$La$_{0.1}$CuO$_{2}$,\ Ni reduces $T_{c}$ much faster than Zn}
\author{C. U. Jung, J. Y. Kim, Min-Seok Park, Heon-Jung Kim, Mun-Seog Kim, and
Sung-Ik Lee\cite{email}}
\address{National Creative Research Initiative Center for Superconductivity and\\
Department of Physics, Pohang University of Science and Technology, Pohang\\
790-784, Republic of Korea }
\date{\today }
\maketitle

\begin{abstract}
We report the effect of substitution for Cu on the $T_{c}$\ of the
electron-doped infinite-layer superconductors Sr$_{0.9}$La$_{0.1}$Cu$_{1-x}$%
{\it R}$_{x}$O$_{2}$ for {\it R} = Zn and Ni. We found that $T_{c}$\ was
nearly constant until $x\sim 0.03$\ for {\it R} = Zn while the
superconductivity was nearly suppressed for $x\sim 0.02$\ with $%
dT_{c}/dx\geq 20$ K/\% for {\it R} = Ni. This behavior is very similar to
that of conventional superconductors. These findings are discussed in terms
of the superconducting gap symmetry in the cuprate superconductors,
including another electron-doped superconductor, (Nd,Ce)$_{2}$CuO$_{4-y}$.
\end{abstract}



\begin{multicols}{2}
\section{Introduction}

Impurity substitution at the Cu site in\ high-$T_{c}$ cuprates\cite
{nt328,nt332} has been considered to be a test probe for the mechanism of
high-temperature superconductivity and the symmetry of the superconducting
order parameter [3-33]. It has also shed light on the recent striking
issue of a normal-state pseudo-gap.\cite{l79p5294,b58p5956,c331p38} The most
important observation was that the non-magnetic Zn ion suppressed $T_{c}$
somewhat more than magnetic ions such as Ni for all kinds of hole-doped
cuprates such as (La,Sr)$_{2}$CuO$_{4}$,\cite
{l79p5294,l77p5421,b42p218,c317p575} YBa$_{2}$Cu$_{3}$O$_{7-\delta }$,\cite
{l79p5294,l77p5421,l76p684,epl46p678,l67p2088,b50p4051,l84p3422} YBa$_{2}$Cu$%
_{4}$O$_{8}$,\cite{b61p4319,b44p10139} and Bi$_{2}$Sr$_{2}$CaCu$_{2}$O$_{8}$%
\cite{b58p5956,c331p38}. The $T_{c}$ reduction rate for Zn substitution was $%
dT_{c}/dx\sim 10$ K/\% and was higher for underdoped compounds than
optimally or overdoped ones. This behavior of $T_{c}$ is strongly contrasted
with that of conventional superconductors, where the reduction of $T_{c}$ is
stronger for magnetic ion impurities, but nearly absent for nonmagnetic ion
impurities. This difference led to the theoretical formulation of an
unconventional pairing mechanism and a symmetry of the order parameter for\
high-temperature superconductor [16-27].

Substitution at the Cu site in ordinary high-$T_{c}$ cuprates with a charge
reservoir block is generally not immune to structural distortion and/or
charge carrier transfer between the charge reservoir block and the
conducting CuO$_{2}$ planes, which could affect $T_{c}$ dramatically.
Especially, one must be very cautious about the oxygen content when
comparing the amounts of $T_{c}$ reduction directly. Another complexity is
due to the existence of several substitutable sites inside a unit-cell.

Electron-doped infinite-layer superconductors (Sr$_{1-x}^{+2}$Ln$_{x}^{+3}$%
)CuO$_{2}$ (Ln = La, Sm, Nd, Gd, etc.)\cite{Nature91,Review}\ have several
incomparable merits for studying the effect of substitution for Cu on $T_{c}$%
. Without a charge reservoir block, it has only the back-bone structure
common to all high-$T_{c}$ cuprates, CuO$_{2}$ planes separated only by a
metallic spacer layer.\cite{Nature88} The structure is robust, and oxygen is
very stoichiometric and stable: buckling of the CuO$_{2}$ plane, O
interstitials and O vacancies were reported to be nearly absent.\cite
{Buckling} The $T_{c}$ has been found to be very robust against
modifications of the structure and changes in the magnetic moment due to
doping various lanthanide ions at the Sr sites.\cite{Ikeda93} Thus, the
substitution\ for Cu in the infinite-layer superconductors has the least
possibility of changing $T_{c}$ via secondary routes, and observations
should reveal a more intrinsic change in the $T_{c}$ of cuprate
superconductors. However, difficulties in synthesizing high-quality sample
has prohibited intense work on these infinite-layer superconductors.\cite
{Jungwork} Moreover, no work on the Cu-substitution effect has been reported.

Recently, we succeeded in synthesizing high-quality (Sr$_{0.9}$La$_{0.1}$)CuO%
$_{2}$.\cite{Jungwork} Here, we report the effect of substitution for Cu on
the $T_{c}$\ of the electron-doped infinite-layer superconductors Sr$_{0.9}$%
La$_{0.1}$Cu$_{1-x}${\it R}$_{x}$O$_{2}$ where {\it R} = Zn and Ni. We found
that $T_{c}$\ was nearly constant until $x\sim 0.03$\ for R = Zn while the
superconductivity was nearly suppressed for $x\sim 0.02$\ with $%
dT_{c}/dx\geq 20$ K/\% for Ni. This feature is similar to those observed in
conventional superconductors.

\section{Experimental}

Starting materials of La$_{2}$O$_{3}$, SrCO$_{3}$, CuO, and ZnO (NiO) with a
nominal composition were calcined at $920\sim 945$ $^{\circ }$C\ for 36
hours with several intermittent grindings. A pelletized precursor sandwiched
between two Ti oxygen-getter slabs was put in a Au or Pt capsule. The
capsule, together with an insulating wall and a graphite-sleeve heater, was
closely packed inside a high-pressure cell made of pyrophillite. Details of
the sintering under high pressure is found elsewhere.\cite{Jungwork} The
masses of the homogenous samples obtained in one batch were larger than $200$
mg. The low-field magnetization was measured in the zero-field-cooled state
by using a SQUID magnetometer (MPMS{\it XL}, Quantum Design) at $10\sim 20$
Oe. The powder X-ray diffraction (XRD) was measured using a RIGAKU X-ray
diffractometer. Energy dispersive spectroscopy (EDS) using an electron probe
microanalyzer and a field emission scanning electron microscope (JSM-6330F,
JEOL) were also used.

\section{Data analysis and discussion}

Figure \ref{XRDMT}(a) shows the X-ray powder diffraction patterns of Sr$%
_{0.9}$La$_{0.1}$Cu$_{1-x}$Zn$_{x}$O$_{2}$, $x=0$, 0.01, 0.03, and Sr$_{0.9}$%
La$_{0.1}$Cu$_{1-x}$Ni$_{x}$O$_{2}$, $x=$ 0.01 and 0.02. The intensity of
each pattern was normalized to the intensity of the (101) peak and offset
vertically for clear comparison. These XRD patterns show that a nearly
single phase with an infinite-layer structure was formed. Peaks
corresponding to Zn oxide, Ni oxide, and La oxide could not be identified
within the resolution. The smaller peaks at $2\theta \sim $ 33.5$^{\circ }$
and 37.5$^{\circ }$ were also found to exist for unsubstituted pristine Sr$%
_{0.9}$La$_{0.1}$CuO$_{2}$ sample with nearly the same diamagnetic signal;
thus they do not correspond to Zn oxide or Ni oxide.

The lattice constants are $a=b=3.928$ (3.950) \AA\ and $c=3.433$ (3.410)\AA\
for insulating SrCuO$_{2}$ (superconducting Sr$_{0.9}$La$_{0.1}$CuO$_{2}$). 
\cite{Buckling,Ikeda93,Jungwork} The expansion of the $a$-axis lattice
constant is known to be due to the transfer of electron carriers to the CuO$%
_{2}$ planes, and the shrinking of the $c$-axis lattice constant is simply
due to the ionic size effect.\cite{Nature91} The lattice constants from the
XRD patterns in Fig. \ref{XRDMT}(a) for Sr$_{0.9}$La$_{0.1}$Cu$_{0.98}$Ni$%
_{0.02}$O$_{2}$ and Sr$_{0.9}$La$_{0.1}$Cu$_{0.97}$Zn$_{0.03}$O$_{2}$\ are $%
a=b=3.943$ \AA\ and $c=3.417$ \AA\ and $a=b=3.950$ \AA\ and $c=3.408$ \AA ,
respectively. The error bars are about 0.003 \AA .

We also examined whether Zn was uniformly distributed within the samples of
Sr$_{0.9}$La$_{0.1}$Cu$_{1-x}$Zn$_{x}$O$_{2}$ with $x=0.03$. The microscopic
composition of the samples was measured over tens of grains, each with a
smaller detecting area of $3\times 3$ $\mu $m$^{2}$. The average diameter of
a grain was about 10 $\mu $m, and the Zn concentration in the grains was $%
3\pm 1$\%. The average Zn concentration was also closer to the nominal
value. Since the entire heating process was done inside a Au capsule, a net
loss of metallic ions is not likely to occur for the high-pressure synthesis
technique. For other samples, the resolution of the EDS was rather
insufficient to determine the stoichiometry.

We measured\ the low-field magnetization and calculated the magnetic
susceptibility, $4\pi \chi (T)$, to determine the effect of substitution at
the Cu site on $T_{c}$. Figure \ref{XRDMT}(b) shows magnetic susceptibility, 
$4\pi \chi (T)$, curves for the above samples. The data for $x=0$\ were from
a previous result.\cite{Jungwork} For Ni substitution, both $T_{c}$ and the
superconducting volume fraction drastically decrease, and superconductivity
nearly vanishes for a 2\% substitution with an average $T_{c}$ reduction
rate of $dT_{c}/dx\geq 20$ K/\%. This behavior was confirmed for several
samples with $x=0.02$. However, for Zn substitution, the change in $T_{c}$
was less than about 2 K until $x\sim 0.03$ where the superconducting volume
fraction became less than about half that of the pristine sample. For
samples with $x\gtrsim 0.03$, growth of singl-phase samples was very
difficult. Though we did not confirm the uniformness of Ni inside the sample
due to the resolution limit of EDS, the result for $dT_{c}/dx$ should be a
lower bound on the real value. The nearly total suppression of
superconductivity in Sr$_{0.9}$La$_{0.1}$Cu$_{0.98}$Ni$_{0.02}$O$_{2}$ seems
not to result from the lattice effect because the lattice constants remain
much closer to those for superconducting Sr$_{0.9}$La$_{0.1}$CuO$_{2}$\ than
those for insulating SrCuO$_{2}$. This means that the reduction of the
electron carrier density in the CuO$_{2}$ plane does not play a dominant
role in killing the superconductivity.

Figure \ref{Tcreduction} shows the reduction rate, $dT_{c}/dx$, for Cu-site
substitution in high-$T_{c}$ cuprates:\ Bi$_{2}$Sr$_{2}$CaCu$_{2}$O$_{8}$, 
\cite{b58p5956,c331p38} (La,Sr)$_{2}$CuO$_{4}$,\cite
{l79p5294,l77p5421,b42p218,c317p575} YBa$_{2}$Cu$_{3}$O$_{7-\delta }$,\cite
{l79p5294,l77p5421,l76p684,epl46p678,l67p2088,b50p4051,l84p3422} (Nd,Ce)$%
_{2} $CuO$_{4}$,\cite{b42p218,reviewnd,b43p10489} and (Sr$_{0.9}$,La$_{0.1}$%
)CuO$_{2}$. The open and the filled symbols represent the cases of Ni and Zn
substitution, respectively.

As noted previously, the $T_{c}$ reduction rate is higher for Zn
substitution in 2-dimensional hole-doped cuprate superconductors, which
occupies the majority of high-$T_{c}$ cuprates. However, this trend was
first reversed for an electron-doped superconductor such as (Nd,Ce)$_{2}$CuO$%
_{4-y}$, which is just one step toward a conventional superconductor in
terms of charge carrier type. Our finding for (Sr$_{0.9}$,La$_{0.1}$)CuO$_{2}
$ seems to be the next step.\ Two kinds of representative n-type cuprates
show similar behaviors for the substitution effect on $T_{c}$, Ni killing
the superconductivity faster than Zn. However the difference is that the
substitution effect in (Sr$_{0.9}$,La$_{0.1}$)CuO$_{2}$ is much closer to
those in conventional superconductors with respect to the exact value of the 
$T_{c}$ reduction rate $dT_{c}/dx$.

Impurity substitution effects in hole-doped cuprates and electron-doped
cuprates have been discussed in terms of the superconducting gap symmetry, 
{\it d}-wave and {\it s}-wave, respectively [16-27]. Phase-sensitive
Josephson tunneling or the presence of a half flux quantum at the center of
the tricrystal ring\cite{nt373} could be a direct test of {\it d}-wave
superconductivity, but would require high-quality thin films, which have not
been feasible for infinite-layer superconductors due to difficulties in film
growth. In addition, the pairing symmetry of the (Nd,Ce)$_{2}$CuO$_{4}$
compound remains to be controversial.\cite{Alff,TsueiPRL00}

Many experimental observations indicate that (Sr$_{0.9}$,La$_{0.1}$)CuO$_{2}$
has properties which are the most similar to those of conventional
superconductors. This compound was reported to have more 3-dimensional
superconductivity with the {\it c}-axis coherence length, even near zero
temperature, being larger than the {\it c}-axis lattice constant.\cite
{MSKimPRL} The undoped antiferromagnetic insulator Ca$_{0.85}$Sr$_{0.15}$CuO$%
_{2}$ has been reported to have more 3-dimensional magnetic coupling and
that material has been reported to have a stronger 3-dimensional character
than other parent insulators of cuprate superconductors, such as YBa$_{2}$Cu$%
_{3}$O$_{6}$, La$_{2}$CuO$_{4}$, and Sr$_{2}$CuO$_{2}$Cl$_{2}$.\cite
{Lombardi96,Vaknin89,Keren93,Pozzi97} For example, an estimate of the ratio
of the out-of-plane to the in-plane coupling constants for Ca$_{0.85}$Sr$%
_{0.15}$CuO$_{2}$\ was two to three orders of magnitude larger than the
corresponding values for YBa$_{2}$Cu$_{3}$O$_{6}$ and La$_{2}$CuO$_{4}$.\cite
{Lombardi96} A very recent observation of scanning tunneling spectra in (Sr$%
_{0.9}$,La$_{0.1}$)CuO$_{2}$ support the existence of an {\it s}-wave gap
with a superconducting gap $\Delta \sim 13$ meV, as well as the absence of
pseudogap.\cite{Yeh} 

\section{ summary}

We found that in electron-doped infinite-layer superconductors Sr$_{0.9}$La$%
_{0.1}$Cu$_{1-x}$R$_{x}$O$_{2}$, substitution of the nonmagnetic Zn ion in
the CuO$_{2}$ plane hardly suppress $T_{c}$ ($dT_{c}/dx\leq 0.5$ K/\% for $%
x\leq 0.03$) while substitution of the magnetic Ni ion kills the
superconductivity at only $x\sim 0.02$ ($dT_{c}/dx\geq 20$ K/\%). This
behavior is similar to that observed for conventional superconductors. This
behavior is also consistent with many recent observations, such as the
existence of {\it s}-wave gap, the stronger 3-dimensionality in
superconducting and antiferromagnetic properties.

\acknowledgments
We greatly appreciate valuable discussions with D. Pavuna, K. Maki, Yunkyu
Bang, N.-C. Yeh, A. V. Balatsky, and M. Sigrist. For the EPMA and SEM
measurements, we are thankful to Mr. Dong Sik Kim at the Department of
Materials Science \& Engineering at Pohang University of Science and
Technology. This work is supported by the Ministry of Science and Technology
of Korea through the Creative Research Initiative Program.

\begin{figure}[tbp]
\caption{(a) X-ray powder diffraction patterns for Sr$_{0.9}$La$_{0.1}$Cu$%
_{1-x}$Zn$_{x}$O$_{2}$ ( $x=0$, 0.01, 0.03) and Sr$_{0.9}$La$_{0.1}$Cu$_{1-x}
$Ni$_{x}$O$_{2}$ ( $x=$ 0.01 and 0.02). The intensity of each pattern was
normalized to the intensity of the (101) peak and offset vertically for
clear comparison. (b) Magnetic susceptibility, $4\protect\pi \protect\chi (T)
$, curves for the samples. }
\label{XRDMT}
\end{figure}

\begin{figure}[tbp]
\caption{$T_{c}$ reduction rate, $dT_{c}/dx$, for Cu-site substitution in
high-$T_{c}$ cuprates; open symbols are for Ni substitution and closed
symbols are for Zn substitution. The squares are for\ Bi$_{2}$Sr$_{2}$CaCu$%
_{2}$O$_{8}$, the circles for (La,Sr)$_{2}$CuO$_{4}$, the up-triangles for
YBa$_{2}$Cu$_{3}$O$_{7-\protect\delta }$, the down-triangles for (Nd,Ce)$_{2}
$CuO$_{4}$, and the diamonds for Sr$_{0.9}$La$_{0.1}$CuO$_{2}$. Note that
non-magnetic Zn hardly suppresses the $T_{c}$ of an infinite-layer
superconductor.}
\label{Tcreduction}
\end{figure}

\end{multicols}

\end{document}